\documentclass{article}
\usepackage[english]{babel}
\usepackage{amssymb}
\usepackage{latexsym}
\usepackage{exscale}
\usepackage{psfig}
\begin{document}

\renewcommand{\thefootnote}{\fnsymbol{footnote}}

\begin{center}
{\bf \Large Entanglement degradation in continuous-variable
quantum teleportation}\footnote{Contribution to ICSSUR VII, Boston,
2001}\\[3ex] 
{\large Dirk-Gunnar Welsch\footnote{E-mail: welsch@tpi.uni-jena.de},
Stefan Scheel, Aleksej V. Chizhov\footnote{Permanent address: Joint
Institute for Nuclear Research, Laboratory of Theoretical Physics,
141980 Dubna, Russia.}}\\[1ex]
Theoretisch-Physikalisches Institut, 
Friedrich-Schiller-Universit\"at Jena, Max-Wien-Platz 1,
D-07743 Jena, Germany\\[3ex]

\noindent
\parbox{.9\textwidth}{
{\large \bf Abstract}\\[1ex]
{\small 
The influence of losses in
the transmission of continuous-variable entangled light
through linear devices such as optical fibers
is studied, with special emphasis on Gaussian states.
Upper bounds on entanglement
and the distance to the set of separable Gaussian states are
calculated. Compared with the distance measure, the bounds
can substantially overestimate the entanglement and thus 
do not show the drastic decrease of entanglement with
increasing mean photon number, as does the distance measure. 
In particular, it shows that losses give rise to entanglement
saturation, which principally limits the amount of information that
can be transferred quantum mechanically in continuous-variable
teleportation. Even for an initially infinitely squeezed two-mode
squeezed vacuum, high-fidelity teleportation is only
possible over distances that are much smaller than the
absorption lengths.  
}
}
\end{center}


\section{Introduction}
\label{sec1}

Entangled quantum states containing more than one photon on
average have been of increasing interest. So, in
continuous-variable quantum teleportation the entangled
state shared by Alice and Bob is commonly assumed to be a
Gaussian EPR-type state such as an infinitely squeezed two-mode
squeezed vacuum (TMSV) \cite{Braunstein98}, which then
represents an entangled macroscopic state. Since entanglement
is a nonclassical property, a strongly squeezed
TMSV may be expected to be very instable; that is, strong
entanglement degradation due to dissipative environments
may be expected. This would of course 
have dramatic consequences for teleportation over reasonable 
distances of an arbitrary quantum state with sufficiently high
fidelity.

The aim of the present paper is to investigate the problem
of entanglement degradation in transmission of a TMSV
through generically lossy optical systems such as fibers,
with special emphasis on the ultimate limits in
continuous-variable quantum teleportation. The
necessarily existing interaction of the light with 
dissipative environments spoil the quantum-state purity,
leaving behind a statistical mixture. Unfortunately, quantification
of entanglement for mixed states in infinite-dimensional Hilbert
spaces is yet impossible in practice. It typically involves
minimizations over infinitely many parameters, as it is the
case for the entropy of formation as well as for the distance
to the set of all separable quantum states measured either
by the relative entropy or Bures' metric \cite{Vedral98}.
It is, however, possible to derive upper
bounds on the entanglement content \cite{Hiroshima00} by using the
convexity property of the relative entropy.

For Gaussian states, however, an upper bound based on the distance
to the set of separable Gaussian states can be derived, which 
is far better than the convexity bound. In particular, the 
entanglement degradation found in this way closely corresponds to the 
reduction of the fidelity of quantum teleportation. In this
way, ultimate limits of quantum information transfer can be
derived, which show that only a small fraction of the amount of
quantum information that would be contained in an infinitely
squeezed TMSV is really available in praxis.


\section{Entanglement degradation}
\label{sec2}

Let us consider the problem of entanglement degradation
in transmission of light prepared in a TMSV state
\begin{equation}
\label{3.1}
|\mbox{TMSV}\rangle
= \exp\!\left(\zeta^\ast\hat{a}_1\hat{a_2}
- \zeta\hat{a}_1^\dagger\hat{a_2}^\dagger \right)|0,0\rangle
= \sqrt{1-|q|^2} \,\sum\limits_{n=0}^\infty (-q)^n |n,n\rangle
\end{equation}
($q$ $\!=$ $\!e^{i\phi}\tanh|\zeta|$, $\phi$ $\!=$ $\!\arg \zeta$)
through absorbing fibers characterized by the transmission
coefficients $T_i$ (\mbox{$i$ $\!=$ $\!1,2$}), which are given by
the Lambert--Beer law of extinction,
\begin{equation}
\label{3.11}
|T_i| = e^{-l_i/l_{{\rm A}i}}
\end{equation}
($l_i$, transmission length; $l_{{\rm A}i}$, absorption length).
The quantum state of the transmitted field can be found
by applying the formalism developed in \cite{Knoll99,Scheel00c}. In the
number basis it reads
\begin{equation}
\label{3.2}
\hat{\varrho} = ( 1-|q|^2)
\sum\limits_{m=0}^\infty \sum\limits_{k,l=0}^\infty
\Big[ K_{k,l,m}
( c_m |m\!+\!k\rangle\langle k| +\mbox{H.c.})
\otimes
( d_m |m\!+\!l\rangle\langle l| +\mbox{H.c.})\Big],
\end{equation}
where
\begin{eqnarray}
\label{3.3}
c_m &=& (-q)^{m/2} T_1^m
\left( 1-{\textstyle\frac{1}{2}}\delta_{m0} \right), \\
d_m &=& (-q)^{m/2} T_2^m
\left( 1-{\textstyle\frac{1}{2}}\delta_{m0} \right),
\end{eqnarray}
and
\begin{eqnarray}
\label{3.4}
\lefteqn{
K_{k,l,m} =
\frac{\big[|q|^2 (1\!-\!|T_1|^2)
(1\!-\!|T_2|^2) \big]^a a! (a\!+\!m)!}
{\sqrt{k!l!(k\!+\!m)!(l\!+\!m)!}(a\!-\!k)!(a\!-\!l)!}
}
\nonumber \\ && \hspace{2ex}
\times
\left( \frac{|T_1|^2}{1-|T_1|^2} \right)^k
\left( \frac{|T_2|^2}{1-|T_2|^2} \right)^l
{}_2F_1\!\left[ {a\!+\!1,\,a\!+\!m\!+\!1
\atop |k\!-\!l|\!+\!1};\,
|q|^2(1\!-\!|T_1|^2)(1\!-\!|T_2|^2) \right]
\end{eqnarray}
[$a$ $\!=$ $\!\max(k,l)$].
The state is a Gaussian state whose Wigner function reads
\begin{equation}
\label{2.9}
W(\alpha_1,\alpha_2) = \frac{4}{\pi ^2 \cal{N}}
   \exp\!\left[- 2 \bigl(C_2|\alpha_1|^2+C_1|\alpha_2|^2  
   + S^\ast\alpha_1\alpha_2 + S\alpha_1^* \alpha_2^* \bigr) \right],
\end{equation}
where 
\begin{equation}
\label{2.10} 
S=\frac{e^{i\varphi }}{\cal{N}}\,
   T_1 T_2
   \sinh|2\zeta|,
\end{equation}
\begin{equation}
\label{2.11}
C_i = \frac{1}{\cal{N}}
   \left[ 1 + |T_i|^2 \left(\cosh|2\zeta |-1\right)
   \right] ,
\end{equation}
\begin{equation} 
{\cal{N}} = 1+
   \left( |T_1|^2 + |T_2|^2 -2 |T_1|^2 |T_2|^2 \right)
   \left(\cosh|2\zeta |-1\right).
\label{2.12}
\end{equation}
In Eqs.~(\ref{3.2}) and (\ref{2.9}), the fibers are assumed to
be in the ground state, which implies restriction to optical
fields whose thermal excitation may be disregarded. Otherwise
additional noise is introduced which enhance the entanglement
degradation.


\subsection{Entanglement estimate by pure state extraction}

In order to quantify the entanglement content
of a quantum state $\hat{\varrho}$, we make use of the relative
entropy measuring the distance of the state to the set
${\cal S}$ of all separable states $\hat{\sigma}$
\cite{Vedral98},
\begin{equation}
\label{2.25}
E(\hat{\varrho})= \min_{\hat{\sigma}\in {\cal S}}
{\rm Tr}\big[ \hat{\varrho}
\big( \ln \hat{\varrho} - \ln \hat{\sigma} \big) \big].
\end{equation}
For pure states this measure reduces to the von Neumann entropy
\begin{equation}
\label{2.10a}
E(\hat{\varrho})
   = S_{1(2)} = -{\rm Tr}\bigl[ \hat{\varrho}_{1(2)} \ln
\hat{\varrho}_{1(2)} \bigr],
\end{equation}
where $\hat{\varrho}_{1(2)}$ denotes the (reduced) output
density operator of mode $1(2)$, which is obtained by
tracing $\hat{\varrho}$ with respect to mode $2(1)$. Thus,
\begin{equation}
\label{3.12}
E\big(|\mbox{TMSV}\rangle\big) =
-\ln \big( 1\!-\!|q|^2 \big) -\frac{|q|^2}{1\!-\!|q|^2} \ln |q|^2 .
\end{equation}

Unfortunately, there is no closed solution of
Eq.~(\ref{2.25}) for arbitrary mixed states. Nevertheless,
upper bounds on the entanglement can be calculated,
using the convexity of the relative entropy. When
\begin{equation}
\label{2.26a}
\hat{\varrho} = \sum_n p_n\hat{\varrho}_n,
\quad \sum_n p_n =1
\end{equation}
($p_n$ $\!\ge$ $\!0$),
then $E(\hat{\varrho})$ satisfies the inequality
\begin{equation}
\label{2.26}
E(\hat{\varrho}) \le \sum_n p_n E(\hat{\varrho}_n).
\end{equation}
If the $E(\hat{\varrho}_n)$ are known, this inequality can be used
to introduce an upper bound. Apart from pure states,
it is known that when the quantum state has the Schmidt form, 
\begin{equation}
\label{2.27}
\hat{\varrho} = \sum\limits_{n,m} C_{n,m}
|\phi_n,\psi_n\rangle\langle \phi_m,\psi_m|,
\end{equation}
then the amount of entanglement measured by the
relative entropy is given by \cite{Rains99,Wu00}
\begin{equation}
\label{2.28}
E(\hat{\varrho})
= -\sum_n C_{n,n} \ln C_{n,n} - S(\hat{\varrho}).
\end{equation}


\begin{figure}[htb]
\begin{center}
\mbox{\psfig{file=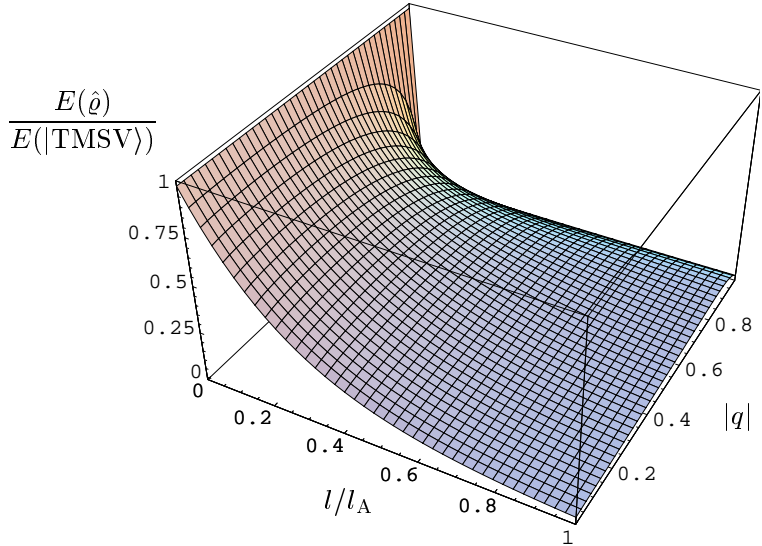,width=10cm,clip=}}
\end{center}
\caption{\label{estimate}
Estimate of the entanglement, Eq.~(\protect\ref{3.15}),
observed after transmission of a TMSV through
absorbing fibers \mbox{($T_1$ $\!=$ $\!T_2$)}    
as a function of the squeezing parameter
\mbox{$|q|$ $\!=$ $\!\sinh|\zeta|$}
and the transmission \mbox{length $l$.}}
\end{figure}

\subsubsection{Extraction of a single pure state}
\label{sec:single}

Since, by Eqs.~(\ref{3.2}) -- (\ref{3.4}), for low squeezing
only a few matrix elements are excited
which were not contained in the original Fock expansion (\ref{3.1}),
we can forget about the entanglement that could be present in the newly
excited elements and treat them as contributions to the separable
states only. The inseparable state relevant for entanglement might
then be estimated to be the pure state  
\begin{equation}
\label{3.13}
\sqrt{1-\lambda}\, |\Psi\rangle 
= \sqrt{\frac{1-|q|^2}{K_{000}}} \sum\limits_{n=0}^\infty
K_{00n} c_n d_n |n,n\rangle.
\end{equation}
It has the properties that only matrix elements 
of the same type as in the input TMSV state occur and
the coefficients of the matrix elements \mbox{$|0,0\rangle$ 
$\!\leftrightarrow$ $\!|n,n\rangle$} are met exactly, i.e.,
\begin{equation}
\label{3.14}
(1-\lambda)\langle 0,0|\Psi\rangle\langle\Psi|n,n\rangle
=\langle 0,0|\hat{\varrho}|n,n\rangle. 
\end{equation} 
In this approximation, the calculation of the entanglement
of the mixed output quantum state 
reduces to the determination of the entanglement of a
pure state \cite{Scheel00c}:
\begin{eqnarray}
\label{3.15}
\lefteqn{
E(\hat{\varrho}) \approx (1-\lambda)\,E(|\Psi\rangle) 
}
\nonumber\\[.5ex]&&\hspace{2ex}
   =\frac{1\!-\!x}{(1\!-\!x)^2\!-\!y}\,
   \ln\!\left[\frac{1\!-\!x}{(1\!-\!x)^2\!-\!y}\right] 
   +\, \frac{(1\!-\!x)\{[y\!+\!(1\!-\!x)^2] 
   \ln (1\!-\!x)\!-\!y \ln y\}}{[y\!-\!(1\!\!-x)^2]^2} \,,
\end{eqnarray}
where
\begin{equation}
\label{3.16}
x = |q|^2 (1-|T_1|^2) (1-|T_2|^2) ,
\end{equation}
\begin{equation}
\label{3.17}
y = |qT_1T_2|^2 .
\end{equation}
In Fig.~\ref{estimate}, the estimate of entanglement as given by
Eq.~(\ref{3.15}) is plotted as a function of the
transmission length and the strength of initial
squeezing for $T_1$ $\!=$ $\!T_2$ $\!=$ $\!T$.


\subsubsection{Upper bound of entanglement}
\label{sec:rains}

\begin{figure}[htb]
\begin{center}
\mbox{\psfig{file=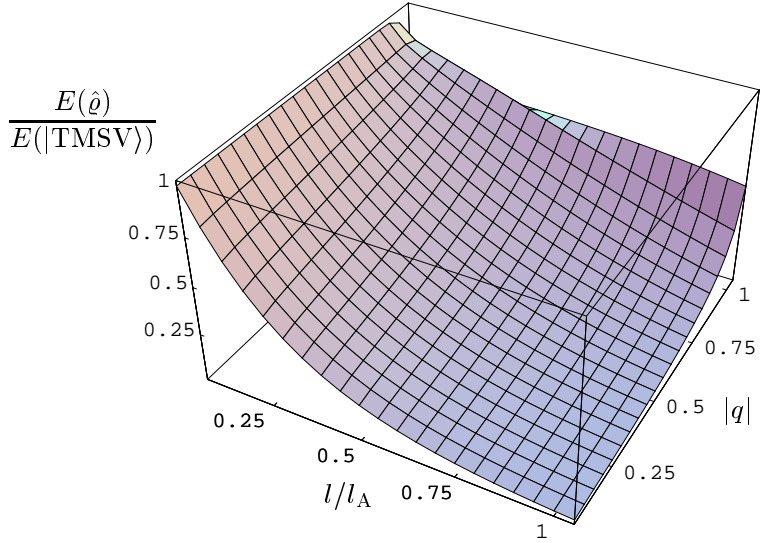,width=10cm,clip=}}
\end{center}
\caption{\label{tmsv_est}
Upper bound on the entanglement degradation of a TMSV
transmitted through absorbing
fibers \mbox{($T_1$ $\!=$ $\!T_2$)} as a function of the
squeezing parameter \mbox{$|q|$ $\!=$ $\!\sinh|\zeta|$}
and the transmission length $l$.
In the numerical calculation, at most 30 photons per mode
have been taken into account which is obviously not sufficient
for higher squeezing when higher photon-number states are
excited.
}
\end{figure}

As already mentioned, the estimate given by Eq.~(\ref{3.15}) is
valid for low squeezing. Higher squeezing amounts to more excited
density matrix elements and Eq.~(\ref{3.15}) might become wrong.
Moreover, we cannot even infer it to be a {\it bound} in any sense
since no inequality has been involved. A possible way out would be to
extract successively more and more pure states from
Eq.~(\ref{3.2}). But instead, let us follow Ref. \cite{Hiroshima00},
writing the density operator (\ref{3.2}) as the convex sum of
density operators in Schmidt decomposition,
\begin{equation}
\label{3.20}
\hat{\varrho}=
\sum\limits_{k,l=0}^\infty \bigg\{
A_{k,l} |k,k\rangle\langle l,l|
+\sum\limits_{m=1}^\infty B_{k,l,m} |k\!+\!m,k\rangle\langle l\!+\!m,l|
+\sum\limits_{m=1}^\infty C_{k,l,m} |k,k\!+\!m\rangle\langle l,l\!+\!m|
\bigg\},
\end{equation}
and applying the inequality (\ref{2.26}) together with
Eq.~(\ref{2.28}). The result is illustrated in Fig.~\ref{tmsv_est}.

{F}rom general arguments one would expect the entanglement to decrease
faster the more squeezing one puts into the TMSV, because stronger
squeezing is equivalent to saying the state is more macroscopically
non-classical and quantum correlations should be destroyed faster. As
an example, one would have to look at the entanglement degradation of
an $n$-photon Bell-type state $|\Psi^\pm_n\rangle$, 
$E[\hat{\varrho}(|\Psi^\pm_n\rangle)]$ $\!\le$ $\!|T|^{2n} \ln 2$
\cite{Scheel00a}. 
Since the transmission coefficient $T$ decreases exponentially
with the transmission length, entanglement decreases even faster.
Note that similar arguments also hold for the destruction of
the interference pattern of a cat-like state
$\sim|\alpha\rangle+|-\alpha\rangle$ when it is transmitted,
e.g., through a beam splitter. It is well known 
that the two peaks (in the $j$th output channel) decay as $|T_{j}|^2$,
whereas the quantum interference decays as
$|T_{j}|^2 \exp[-2|\alpha|^2(1$ $\!-$ $\!|T_{j}|^2)]$. 

The upper bound on the entanglement as calculated above
seems to suggest that the entanglement degradation
is simply exponential with the transmission length for
essentially all (initial) squeezing parameters, which would
make the TMSV a good candidate for a robust entangled quantum
state. But this is a fallacy. The higher the
squeezing, the more density matrix elements are excited, and the more
terms appear, according to Eq.~(\ref{3.20}), in the
convex sum (\ref{2.26}). Equivalently, more and
more separable states are mixed into the full quantum state.
By that, the inequality gets more inadequate.
We would thus conclude that
the upper bound proposed in \cite{Hiroshima00} is insufficient.


\subsection{Distance to separable Gaussian states}
\label{sec:distance}

The methods of computing entanglement estimates and bounds as
considered in the preceding sections are  based on Fock-state
expansions. In practice they are typically restricted to situations
where only a few quanta of the overall system (consisting of
the field and the device) are excited, 
otherwise the calculation even of the matrix elements
becomes arduous. Here we will focus on another way
of computing the relative entropy, which will also enable us to
give an essentially better bound on the entanglement.  

Since it is close to impossible to compute the distance of a Gaussian
state to the set of {\em all} separable states we restrict ourselves
to separable {\em Gaussian} states. A quantum state is commonly called
Gaussian if its Wigner function is Gaussian. The density operator of
such a state can be written in exponential form of
\begin{equation}
\label{4.2}
\hat{\sigma} = {\cal N}_\sigma
\exp\!\left[ -\left(\hat{a}^\dagger \, \hat{a}\right) {\bf M}_\sigma
{\hat{a} \choose \hat{a}^\dagger } \right],
\end{equation}
where ${\bf M}_\sigma$ is a Hermitian matrix, which can be assumed
to give a symmetrically ordered density operator, and ${\cal N}_\sigma$
is a suitable normalization factor. Here and in the following
we restrict ourselves to Gaussian states with zero mean.
Since coherent displacements, being local unitary transformations,
do not influence entanglement, they can be disregarded. 

The relative entropy (\ref{2.25}) can now be written as
\begin{equation}
\label{4.3}
E\big(\hat{\varrho}\big) = 
   {\rm Tr}\left(\hat{\varrho} \ln \hat{\varrho}\right)
   - \min_{\hat{\sigma}\in{\cal S}} {\rm Tr}
   \left\{
   \hat{\varrho}
   \left[
   \ln {\cal N}_\sigma
   - \left(\hat{a}^\dagger \, \hat{a}\right) {\bf M}_\sigma
   {\hat{a} \choose \hat{a}^\dagger }
   \right]
   \right\}.
\end{equation}
Since we have chosen the density operator $\hat{\sigma}$ to be
symmetrically ordered, the last term in Eq.~(\ref{4.3}) is nothing but 
a sum of (weighted) symmetrically ordered expectation values
$\langle \hat{a}^{\dagger m} \hat{a}^n \rangle_{s=0}$
($m$ $\!+$ $\!n$ $\!=$ $\!2$). Equation (\ref{4.3})
can equivalently be given in terms of
the matrix ${\bf D}_\varrho$ in the exponential of the
characteristic function of the Wigner function as
\begin{equation}
\label{4.4}
E\big(\hat{\varrho}\big) = {\rm Tr}\left( \hat{\varrho} \ln
   \hat{\varrho} \right)
   + \min_{\hat{\sigma}\in{\cal S}}
   \!\left[ \textstyle\frac{1}{2}
   {\rm Tr}\left({\bf M}_\sigma {\bf D}_\varrho\right)
   -\ln {\cal N}_\sigma \right] . 
\end{equation}

The matrix ${\bf D}_\varrho$ is unitarily equivalent to the variance
matrix ${\bf V}$. For a Gaussian distribution with zero mean the 
elements of the variance matrix are defined by
the (symmetrically ordered) correlations of the quadrature
components $\hat{x}_1,\hat{p}_1,\hat{x}_2,\hat{p}_2$.
The variance matrix of any Gaussian state can be brought to
the generic form 
\begin{equation}
\label{4.6a}
{\bf V} = \left(
\begin{array}{cccc}
x&0&z_1&0\\0&x&0&z_2\\z_1&0&y&0\\0&z_2&0&y
\end{array} \right)
\end{equation}
by local Sp(2,$\mathbb{R}$)$\otimes$Sp(2,$\mathbb{R}$) transformations
\cite{Simon00}, so that we can restrict our attention to that case.
Separability requires that \cite{Simon00,Duan00a} 
\begin{equation}
\label{4.11}
4(xy-z_1^2)(xy-z_2^2) \ge (x^2+y^2) +2|z_1z_2|
   -\textstyle\frac{1}{4} \,.
\end{equation}
{F}rom Eq.~(\ref{2.9}) [together with Eqs.~(\ref{2.10}) and
(\ref{2.11})] it follows that we may set
\begin{equation}
\label{4.7}
x = {\textstyle\frac{1}{2}}{\cal N}C_1,
\quad
y = {\textstyle\frac{1}{2}}{\cal N}C_2,
\end{equation}
\begin{equation}
\label{4.8}
z_1 = - z_2 = {\textstyle\frac{1}{2}}{\cal N} |S|.
\end{equation}
Combining Eqs.~(\ref{4.11}) -- (\ref{4.8}), it is not
difficult to prove that the boundary between separability and
inseparability is reached for \mbox{$T_1$ $\!=$ $\!T_2$ $\!=$
$\!0$}; that is, for infinite transmission length
\cite{Scheel00c,Duan00a,Lee00}. Clearly, this tells us
nothing about the entanglement degradation and the
amount of entanglement that is really available for chosen
transmission lengths.

\begin{figure}[htb]
\begin{center}
\mbox{\psfig{file=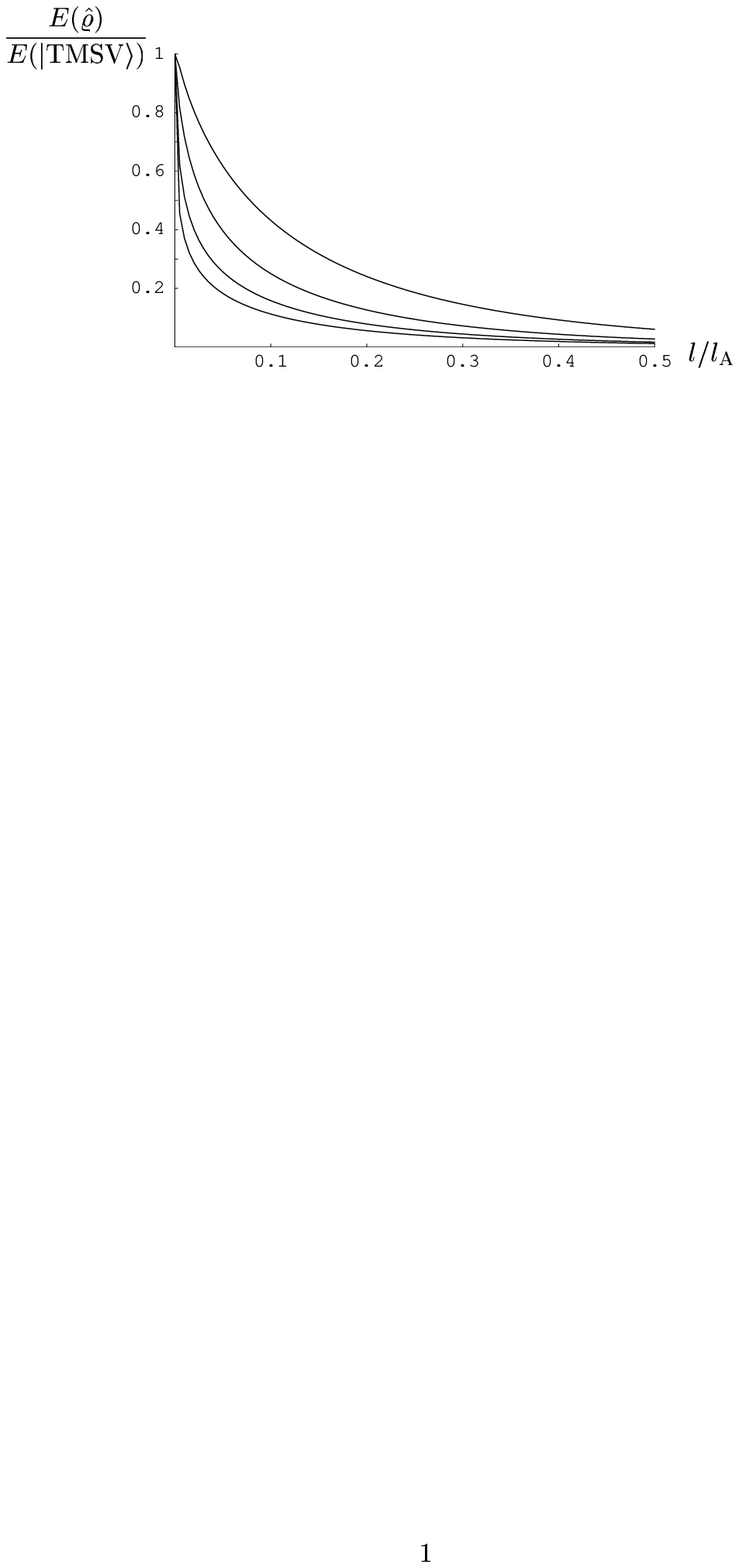,width=10cm,clip=}}
\end{center}
\caption{\label{distrel}
Entanglement
degradation of a TMSV transmitted through absorbing
fibers \mbox{($T_1$ $\!=$ $\!T_2$)} as a function of the
transmission length $l$ for the (initial) mean photon numbers
\mbox{$\bar{n}$ $\!=$ $1$} (\mbox{$|q|$ $\!\simeq$ $\!0.7071$})
(topmost curve),
\mbox{$\bar{n}$ $\!=$ $10$} (\mbox{$|q|$ $\!\simeq$ $\!0.9535$}),
\mbox{$\bar{n}$ $\!=$ $10^2$} (\mbox{$|q|$ $\!\simeq$ $\!0.9950$}), and
\mbox{$\bar{n}$ $\!=$ $10^3$} (\mbox{$|q|$ $\!\simeq$ $\!0.9995$})
(lowest curve).
}
\end{figure}

In order to obtain a measure of the
entanglement degradation, we therefore compute the distance
of the quantum state of the transmitted light to the set of
all Gaussian states satisfying the {\em equality} in (\ref{4.11}),
since they just represent the boundary between separability
and inseparability. These states are completely specified by only three
real parameters [one of the parameters in the equality in (\ref{4.11})
can be computed by the other three]. With regard to Eq.~(\ref{4.4}),
minimization is thus only performed in a three-dimensional
parameter space.
Results of the numerical analysis are shown in
Fig.~\ref{distrel}. It is clearly seen that the entanglement
content (relative to the entanglement in the initial TMSV)
decreases noticeably faster for larger squeezing, or equivalently,
for higher mean photon number [the relation between the
mean photon number $\bar{n}$ and the squeezing parameters
being $\bar{n}$ $\!=$ $\!\sinh^2|\zeta|$ $\!=$
$\!|q|^2/(1$ $\!-$ $\!|q|^2)$].

\subsection{Comparison of the methods}
\label{sec:comparison}

\begin{figure}[htb]
\begin{center}
\mbox{\psfig{file=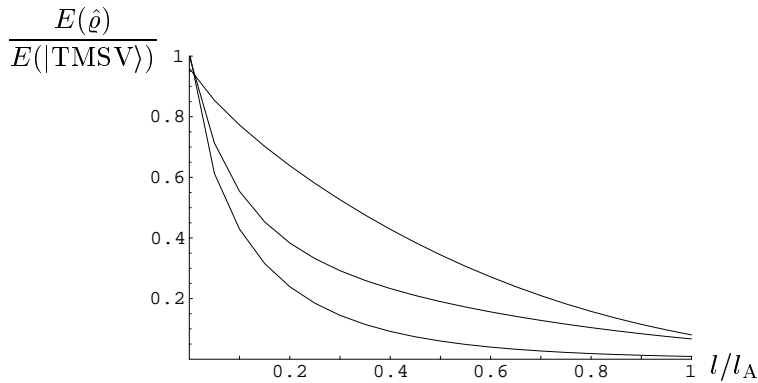,width=10cm,clip=}}
\end{center}
\caption{\label{abstandsvergleich} Comparison of the upper bound on
entanglement (upper curve) according to Fig.~\protect\ref{tmsv_est},
the entanglement estimate (middle curve) according to
Fig.~\protect\ref{estimate}, and the distance measure
(lower curve) according to Fig.~\protect\ref{distrel}
for the mean photon number \mbox{$\bar{n}$ $\!=$ $1$}
(\mbox{$|q|$ $\!\simeq$ $\!0.7071$}).
}
\end{figure}

In Fig.~\ref{abstandsvergleich} the entanglement degradation
as calculated in Section \ref{sec:distance} is compared with the
estimate obtained in Section \ref{sec:single} and the bound
obtained in Section \ref{sec:rains}.
The figure reveals
that the distance of the output state to the separable Gaussian
states (lower curve) is much smaller than it might be expected
from the bound on the entanglement (upper curve)
calculated by applying the inequality (\ref{2.26}) [together with
Eqs.~(\ref{2.28})] to Eq.~(\ref{3.20}), as well as
the estimate (middle curve) derived by extracting a 
single pure state according to Eq.~(\ref{3.15}).
Note that the entanglement of the single pure state (\ref{3.13})
comes closest to the distance of the actual state to the separable
Gaussian states, whereas
the convex sum (\ref{3.20}) of density operators in Schmidt
decomposition can give much higher values.
Both methods, however, overestimate the entanglement.
Since with increasing mean photon number
the convex sum contains more and more terms, the bound gets worse
[and substantially slower on the computer, whereas computation of the
distance measure (\ref{4.4}) does not depend on it].

Thus, in our view, the distance to the separable Gaussian states
should be the measure of choice for determining the entanglement
degradation of entangled Gaussian states. Nevertheless, it should be
pointed out that the distance to separable Gaussian states has been
considered and not the distance to all separable states. 
We have no proof yet, that there does not exist an inseparable
non-Gaussian state which is closer than the closest Gaussian state.


\section{Quantum teleportation}
\label{sec3}

\begin{figure}[htb]
\begin{center}
\mbox{\psfig{file=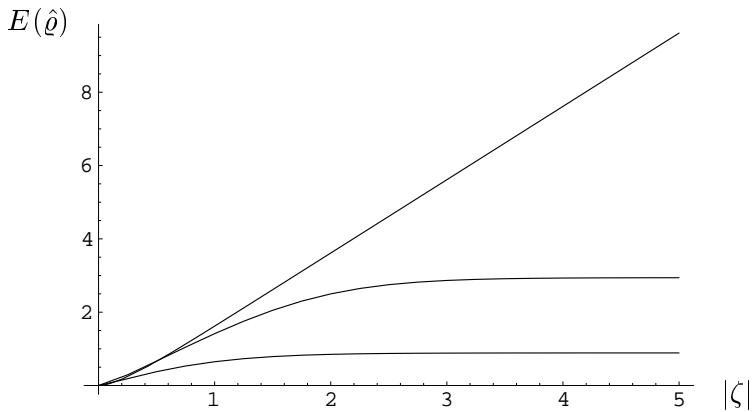,width=10cm}}
\end{center}
\caption{\label{laenge}
Available entanglement after transmission of a TMSV
through absorbing fibers \mbox{($T_1$ $\!=$ $\!T_2$)}
as a function of the squeezing parameter $|\zeta|$ for various
transmission lengths $l$ [%
\mbox{$l$ $\!=$ $\!0$} (topmost curve),
\mbox{$l$ $\!=$ $\!10^{-2}\,l_A$} (middle curve),
\mbox{$l$ $\!=$ $\!10^{-1}\,l_A$} (lowest curve)].
For \mbox{$|\zeta|$ $\!<$ $\!0.5$} and \mbox{$l/l_A$ $\!<$ $\!10^{-2}$},
the numerical accuracy of the values of
$E(\hat{\varrho})$ decreases due to the
low accuracy in the eigenvector computation. 
}
\end{figure}

It is very instructive to know how much entanglement
is available after transmission of the TMSV through the fibers.
Examples of the (maximally) available entanglement 
for different transmission lengths are shown in Fig.~\ref{laenge}.
One observes that a chosen transmission length
allows only for transport of a certain amount of
entanglement. The saturation value, which is quite independent
of the value of the input entanglement,
drastically decreases with increasing transmission length
(compare the upper curve with the two lower curves in the figure).   
This has dramatic consequences for
applications in quantum information processing such as
continuous-variable teleportation, where
a highly squeezed TMSV is required in order
to teleport an arbitrary quantum state with sufficiently
high fidelity. Even if the input TMSV would be
infinitely squeezed, the available (low) saturation value of
entanglement principally prevents one from high-fidelity teleportation
of {\em arbitrary} quantum states over {\em finite} distances.

\subsection{The teleported state}
\label{sec3.1}

\begin{figure}[htb]
\begin{center}
\mbox{\psfig{file=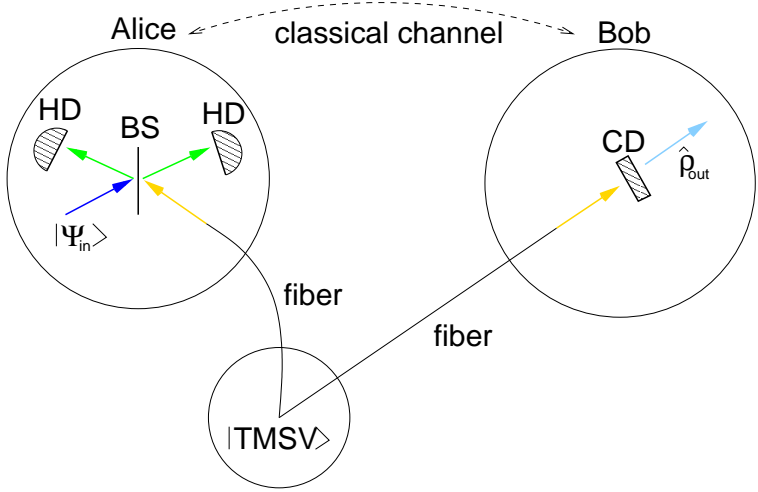,width=10cm}}
\end{center}
\caption{\label{scheme}
Scheme of teleportation ({\textsf BS}, beam splitter;
{\textsf HD}, homodyne detection; {\textsf HD}, coherent
displacement).
}
\end{figure}

Let us briefly repeat the main stages of teleportation
(Fig.~\ref{scheme}).
If $W_{\rm in}(\gamma)$ is the Wigner function of the
signal-mode quantum state that is desired to be teleported
and $W(\alpha,\beta)$ is, according to Eq.~(\ref{2.9})
(\mbox{$\alpha_1$ $\!=$ $\!\alpha$}, \mbox{$\alpha_2$ $\!=$ $\!\beta$}),
the Wigner function of the entangled state that is
shared by Alice and Bob, 
the Wigner function of the (three-mode) overall system then reads
\begin{equation}
\label{2.1} 
W(\gamma,\alpha ,\beta ) = W_{\rm in}(\gamma)\,W(\alpha,\beta).
\end{equation}
After combination of the signal mode and Alice's mode of the
entangled two-mode system through a 50\%:50\% (lossless) beam
splitter the Wigner function changes to
\begin{equation}
\label{2.2} 
W(\mu ,\nu ,\beta )
   = W_{\rm in}\!\left( \frac{\mu -\nu }{\sqrt{2}} \right)
   W\!\left( \frac{\mu +\nu }{\sqrt{2}}, \beta \right).
\end{equation}

Measurement of the real part of $\mu $, $\mu _{\rm R}$,
and the imaginary part of $\nu $, $\nu _{\rm I}$,
then prepares Bob's mode in a quantum state whose
Wigner function is given by
\begin{equation}
\label{2.3}
W(\beta |\mu _{\rm R},\nu _{\rm I}) = \frac{1}{P(\mu _{\rm R},\nu _{\rm I})}
   \int d\nu _{\rm R} \int d\mu _{\rm I} \,
   W_{\rm in}\!\left( \frac{\mu -\nu }{\sqrt{2}} \right)
   W\!\left( \frac{\mu +\nu }{\sqrt{2}}, \beta \right) ,
\end{equation}
where
\begin{equation}
\label{2.4}
P(\mu _{\rm R},\nu _{\rm I})
   = \int d\nu _{\rm R} \int d\mu _{\rm I} \int d^2\beta
   \, W(\mu ,\nu ,\beta )
\end{equation}    
is the probability
density of measuring $\mu _{\rm R}$ and $\nu _{\rm I}$.
Introducuing the complex variables 
\begin{equation}
\label{2.5}
\gamma = \left(\mu - \nu \right)/\sqrt{2},
\quad
\gamma ' =\sqrt{2} \left( \mu _{\rm R} - i \nu _{\rm I} \right),  
\end{equation}
we may rewrite Eq.~(\ref{2.3}) as
\begin{equation}
\label{2.6}
W(\beta |\gamma ')
   = \frac{1}{P(\gamma ')} 
   \int d^2\gamma \,W_{\rm in}(\gamma ) \,
   W({\gamma '}^{*} \!-\! {\gamma }^{*}, \beta ) 
\end{equation} 
[$P(\mu_{\rm R},\nu_{\rm I})/2$ $\!\to$ $\!P(\gamma')$].

Depending upon the result of Alice's measurement,
Bob now coherently displaces the quantum state of his mode
in order to generate a quantum state whose  Wigner function is 
\mbox{$W(\beta$ $\!-$ $\!\Delta(\gamma')|\gamma ')$}.
If we are not interested
in the one or the other measurement result, we may average
over all measurement results to obtain the teleported
quantum state on average:  
\begin{eqnarray}
\label{2.7}
W_{\rm out}(\beta) & = &
   \int d^2\gamma'\, P(\gamma') 
\,
W\!\left(\beta  - \Delta (\gamma ')|\gamma'\right)
\nonumber\\[.5ex]
& = &\int d^2\gamma \,W_{\rm in}(\gamma ) \!\int d^2 \gamma'
   \,W\!\left({\gamma '}^{*} \!-\! {\gamma }^{*},
   \beta \!-\! \Delta (\gamma ')\right) .
\end{eqnarray}

Let us assume that the quantum state to be teleported is
pure, \mbox{$\hat{\varrho}_{\rm in}$ $\!=$
$\!|\psi _{\rm in}\rangle \langle \psi _{\rm in}|$}.
A measure of how close to it is the (mixed) output quantum state
$\hat{\varrho}_{\rm out}$ is the teleportation fidelity
\begin{equation}
\label{2.15}
F
= \langle \psi_{\rm in}| \hat{\varrho}_{\rm out} |\psi _{\rm in}\rangle \,,
\end{equation}
which can be rewritten as the overlap of the Wigner functions:  
\begin{equation}
\label{2.18}
F = \pi \int d^2 \beta\, W_{\rm in} (\beta ) W_{\rm out} (\beta ).
\end{equation}
Perfect teleportation implies unit fidelity;
that is, perfect overlap of the Wigner functions of the
input and the output quantum state. Clearly, losses
prevent one from realizing this case, so that the really
observed fidelity is always less than unity. Thus,
the task is to choose the scheme-inherent parameters
such that the fidelity is maximized. 


\subsection{Choice of the displacement}
\label{sec3.2}

An important parameter that must be specified is the displacement
\mbox{$\beta$ $\!\to$ $\!\beta$ $\!-$ $\!\Delta (\gamma ' )$},
which has to be performed by Bob after Alice's
measurement. For this purpose, we substitute Eq.~(\ref{2.9})
into Eq.~(\ref{2.6}) to obtain, on using the relation
\mbox{$C_1C_2$ $\!-$ $\!|S|^2$ $\!=$ $\!{\cal N}^{-1}$},
\begin{eqnarray}
\label{2.19}
\lefteqn{
W(\beta |\gamma ') = \frac{1}{P(\gamma ')}
   \frac{2}{\pi {C_2\cal{N}}}\,
   \exp\!\left( - \frac{2}{C_2\cal{N}} |\beta |^2 \right)
}
\nonumber\\[.5ex]&&\hspace{2ex}\times    
   \int d^2\gamma\, \frac{2C_2}{\pi }\,
   \exp\!\left( \!-2C_2\left| \gamma ' \!-\! \gamma 
   \!+\!\frac{S^*}{C_2} \beta \right| ^2 \right)
   W_{\rm in} (\gamma ). 
\end{eqnarray}
Here and in the following we restrict our attention to optical fields
whose thermal excitation may be disregarded
(\mbox{$n_{{\rm th}\,i}$ $\!\approx$ $\!0$}).
{F}rom Eqs.~(\ref{2.11}) and (\ref{2.12}) it follows that,
for not too small values of the (initial) squeezing parameter
$|\zeta|$, the variance of the Gaussian in the first line
of Eq.~(\ref{2.19}), $C_2{\cal{N}}/4$,
increases with $|\zeta|$ as $e^{2|\zeta|}|T_2|^2/8$,
whereas the variance of the Gaussian in the integral in the
second line, $1/(4C_2)$, rapidly approaches the (finite) limit
($T_2$ $\!\not=$ $\!0$)
\begin{equation}
\label{2.20}
\sigma _{\infty}= \lim_{|\zeta|\to\infty} \frac{1}{4C_2}
   = \frac{|T_1|^2 + |T_2|^2 - 2 |T_1 T_2|^2}{4 |T_2|^2}\,. 
\end{equation}
Thus, Bob's mode is prepared (after Alice's
measurement) in a quantum state that is obtained,
roughly speaking, from the input quantum state by
shifting the Wigner function according to \mbox{$\gamma$
$\!\to$ $\!\gamma'$ $\!+$ $\!\beta S^*/C_2$} and smearing
it over an area whose linear extension is given by
$2\sqrt{\sigma_\infty}$. It is therefore expected that the
best what Bob can do is to perform a displacement with  
\begin{equation}
\label{asdis}
\Delta (\gamma')
= e^{i\tilde{\varphi} } \lambda \gamma',
\end{equation}
where $\tilde{\varphi}$ $\!=$
$\!\varphi$ $\!+$ $\!\arg T_1$ $\!+$ $\arg T_2$ and
\begin{equation}
\label{lambda}
\lambda  =
\lim_{|\zeta|\to\infty} \frac{C_2}{|S|}
= \left| \frac{T_2}{T_1} \right|.
\end{equation}
Substitution of the expression (\ref{asdis})
into Eq.~(\ref{2.7}) yields
\begin{equation}
\label{wtel}
W_{\rm out}(\beta e^{i\tilde{\varphi}})
= \frac{1}{2\pi\sigma\lambda^2}
\int d^2 \gamma \, W_{\rm in}(\gamma )
\exp\!\left(-\frac{|\gamma - \beta/\lambda|^2}{2\sigma }
\right)\!,
\end{equation}
where
\begin{equation}
\label{sigma}
\sigma
= \frac{\cal{N}}{4 \lambda ^2}
\left(C_2 + \lambda ^2 C_1 - 2\lambda |S| \right).
\end{equation}
Note that $\lim_{|\zeta|\to\infty}\sigma$ $\!=$ $\!\sigma_\infty$.

Clearly, even for arbitrarily large 
squeezing, i.e, \mbox{$|\zeta|$ $\!\to$ $\infty$}, and thus 
arbitrarily large entanglement, the input quantum state
cannot be scanned precisely due to the unavoidable losses, which
drastically reduce, in agreement with the entanglement
degradation calculated in Section \ref{sec2},
the amount of information that can be
transferred nonclassically from Alice to Bob.
Let $\delta_W$ be a measure of the (smallest) scale on which
the Wigner function of the signal-mode state,
$W_{\rm in}(\gamma)$, typically changes. Teleportation
then requires, apart from the scaling by $\lambda$, that
the condition
\begin{equation}
\label{condition}
\sigma _{\infty} \ll \delta _W^2
\end{equation}
is satisfied. Otherwise, essential information about 
the finer points of the quantum state are lost.
For given $\delta _W$, the condition (\ref{condition}) can be
used in order to determine the ultimate limits of teleportation,
such as the maximally possible distance between Alice and
Bob. In this context, the question of the optimal position
of the source of the TMSV arises.
Needless to say, that all the results
are highly state-dependent. Let us study the problem for
squeezed states and Fock states in more detail.


\subsection{Squeezed states}
\label{sec3.3}

The Wigner function of a squeezed coherent state can be given
in the form of
\begin{equation}
\label{wsqin}
W_{\rm in}(\gamma ) =\frac{N_{\rm in}}{\pi }
\,\exp\!\left[ - A_{\rm in}|\gamma |^2
- B_{\rm in}\left(\gamma ^2 +  \gamma ^{\ast 2}\right)
+ C_{\rm in}^\ast\gamma  + C_{\rm in}\gamma^\ast\right],
\end{equation}
where
\begin{equation}
\label{wsqin1}
N_{\rm in} = 2 \exp\!\left[-2 |\alpha _0|^2\cosh\!\left(2\zeta _0\right) 
- \left(\alpha _0^2 +\alpha _0^{*2} \right)
\sinh\!\left( 2\zeta _0\right) \right],
\end{equation}
\begin{equation}
\label{wsqin2}
A_{\rm in} = 2\cosh(2\zeta _0), 
\end{equation}
\begin{equation}
\label{wsqin3}
B_{\rm in} = \sinh(2\zeta _0), 
\end{equation}
\begin{equation}
\label{wsqin4}
C_{\rm in} = 2 \left[ \alpha _0 \cosh \!\left(2\zeta _0\right) 
+ \alpha _0^* \sinh \!\left(2\zeta _0\right) \right].
\end{equation}
Here, $\alpha_0$ is the coherent amplitude and $\zeta_0$ is the
squeezing parameter which is chosen to be real.
Substituting Eq.~(\ref{wsqin}) [together with Eqs.~(\ref{wsqin1}) --
(\ref{wsqin4})] into Eq.~(\ref{wtel}), we derive 
\begin{equation}
\label{wsqout}
W_{\rm out}(\beta ) =\frac{N_{\rm out}}{\pi }
\,\exp\!\left[ - A_{\rm out}|\beta |^2
- B_{\rm out}\left(\beta ^2 +  \beta ^{\ast 2}\right)
+ C_{\rm out}^\ast\beta  + C_{\rm out}\beta^\ast\right],
\end{equation}
where
\begin{equation}
\label{wsqout1}
N_{\rm out} =
\frac{2 
\exp\!\left\{ - \displaystyle\frac{2 |\alpha _0|^2
\left[ \cosh \!\left(2\zeta _0\right) + 4\sigma \right] +
\left( \alpha _0^2 + \alpha _0^{*2} \right) \sinh \!\left(2\zeta _0\right) }
{1+ 8 \sigma \cosh \!\left(2\zeta _0\right) + 16 \sigma ^2} \right\}  }
{\lambda ^2 \sqrt{1+8\sigma\cosh \!\left(2\zeta _0\right)
+ 16\sigma ^2}}\,,
\end{equation}
\begin{equation}
\label{wsqout2}
A_{\rm out} = \frac{2\left[\cosh \!\left(2\zeta _0\right)
+ 4\sigma\right]}{\lambda ^2
\left[1+8\sigma\cosh\!\left( 2\zeta _0\right) + 16\sigma ^2 \right]}\,,
\end{equation}
\begin{equation}
\label{wsqout3}
B_{\rm out} = \frac{\sinh \!\left(2\zeta _0\right)}{\lambda ^2 
\left[1+8\sigma\cosh \!\left(2\zeta _0\right) + 16\sigma ^2 \right]}\,,
\end{equation}
\begin{equation}
\label{wsqout4}
C_{\rm out} = 
2\frac{ \alpha _0 \left[ \cosh \!\left(2\zeta _0\right) + 4\sigma \right] +
\alpha _0^* \sinh \!\left(2\zeta _0\right) }
{\lambda \left[ 1+ 8 \sigma \cosh \!\left(2\zeta _0\right)
+ 16 \sigma ^2 \right] }
\end{equation}
($\tilde{\varphi}$ $\!=$ $\!0$). Combining Eqs.~(\ref{2.18}),
(\ref{wsqin}) -- (\ref{wsqin4}), and (\ref{wsqout}) -- (\ref{wsqout4}),
we arrive at the following
expression for the fidelity
\begin{equation}
\label{fesq}
F \equiv F(\zeta _0,\alpha _0) =
F(\zeta _0)
\exp\!
\left\{ -\frac{(1-\lambda )^2}{2}
\left[
\frac{ \left( \alpha _0 + \alpha _0^{*} \right) ^2 
e^{2\zeta _0} }{1+\lambda ^2
\left(1 + 4 e^{2\zeta _0} \sigma \right)}
- \frac{ \left( \alpha _0 - \alpha _0^{*} \right) ^2 }
{\left(1 + \lambda ^2 \right) e^{2\zeta _0} 
+ 4\lambda ^2 \sigma } 
\right] \right\}
,
\end{equation}
where
\begin{equation}
\label{fesqv}
F(\zeta _0) =
\frac{2}{\sqrt{1 \!+\! 2 \lambda ^2 \!+\! \lambda ^4 \left(1
\!+\! 16 \sigma ^2 \right) \!+\! 8 \lambda ^2
\left(1 \!+\! \lambda ^2 \right) \sigma \cosh\!\left( 2\zeta _0\right)}} 
\end{equation}
is the fidelity for teleporting the squeezed vacuum.

{F}rom Eq.~(\ref{fesq}) it is seen that the dependence on
$\alpha _0$ of $F$ vanishes for \mbox{$\lambda$ $\!=$ $\!1$}.
Thus, the fidelity of teleportation of a squeezed coherent state
can only depend on the coherent amplitude for an asymmetrical
equipment (i.e., \mbox{$|T_1|$ $\!\neq$ $\!|T_2|$}).
In this case, the fidelity exponentially decreases with
increasing coherent amplitude.
For stronger squeezing of the signal mode, the effect is more
pronounced for amplitude squeezing [first term in the
square brackets in the exponential in Eq.~(\ref{fesq})]
than for phase squeezing (second term in the square brackets). 

In the case of a squeezed state, the characteristic scale
$\delta_W$ in the inequality (\ref{condition}) is of the order of
magnitude of the small semi-axis of the squeezing ellipse,
\begin{equation}
\label{3.13*}
\delta_W \sim
e^{-|\zeta _0|}.
\end{equation}
For \mbox{$|T_1|$ $\!\approx$ $\!|T_2|$ $\!=$ $\!|T|$},
from Eqs.~(\ref{2.20})
and (\ref{3.13*}) it then follows that the condition
(\ref{2.25}) for high-fidelity teleportation
corresponds to
\begin{equation}
\label{3.14*}
1 - |T|^2 \ll  e^{-2|\zeta_0|},
\end{equation}
that is,
\begin{equation}
\label{3.15*}
\frac{l}{l_{\rm A}} \ll
\ln\!\left(1 - e^{-2|\zeta_0|} \right)^{-\frac{1}{2}}.
\end{equation}
Thus, for large values of the squeezing parameter $|\zeta_0|$, the
largest teleportation distance that is possible, $l_{\rm T}$,
scales as
\begin{equation}
\label{3.16*}
l_{\rm T} \sim l_{\rm A}\, e^{-2|\zeta_0|}.
\end{equation}    


\subsection{Fock states}
\label{sec3.4}

Let us consider the case when an $N$-photon Fock state
is desired to be teleported. The input Wigner function then reads
\begin{equation}
\label{wfock}
W_{\rm in}(\gamma ) = (-1)^N \frac{2}{\pi} \,
\exp\!\left( -2|\gamma |^2\right)
{\rm L}_N\!\left(4|\gamma |^2\right)
\end{equation}
[${\rm L}_N(x)$, Laguerre polynomial]. We substitute this expression
into Eq.~(\ref{wtel}) and derive the Wigner function of the
teleported state as
\begin{equation}
\label{wfk}
W_{\rm out}(\beta ) = \frac{2}{\pi \lambda ^2}
\frac{(4\sigma - 1)^N}{(4\sigma + 1)^{N+1}}
\exp\!\left[ -\frac{\displaystyle 2|\beta |^2}
{\displaystyle \lambda ^2 (4\sigma + 1)} \right]
\,{\rm L}_N\!\left[ -\frac{4|\beta |^2}
{\lambda ^2 (16\sigma ^2 - 1)} \right].
\end{equation}
Now we combine Eqs.~(\ref{2.18}), (\ref{wfock}), and
(\ref{wfk}) to obtain the teleportation fidelity
\begin{equation}
\label{fidfock}
F \equiv F_{\rm N} = 2 \frac{[\lambda ^2(4\sigma - 1)-1]^N}
{[\lambda ^2(4\sigma + 1)+1]^{N+1}}
\ {\rm P}_N\!\left\{ 1 + \frac{8\lambda ^2}{[\lambda ^2(4\sigma +1)+1]
[\lambda ^2(4\sigma -1)-1]} \right\} 
\end{equation}
[${\rm P}_N(x)$, Legendre polynomial].

{F}rom inspection of Eq.~(\ref{wfock}) it is clear that  
the characteristic scale $\delta_W$ in the inequality
(\ref{2.25}) may be assumed to be of the order of magnitude
of the (difference of two neighbouring) roots of the
Laguerre polynominal ${\rm L}_N(x)$,  
which for large $N$ ({$N$ $\!\gtrsim$ $\!3$}) behaves like $N^{-1}$, thus
\begin{equation}
\label{restrN0}
\delta_{W} \sim \frac{1}{\sqrt{N}}\,.
\end{equation}
Assuming again \mbox{$|T_1|$ $\!\approx$ $\!|T_2|$ $\!=$ $\!|T|$},
the condition (\ref{2.25}) together with Eq.~(\ref{2.20})
and $\delta_W^2$ according to Eq.~(\ref{restrN0}) gives 
\begin{equation}
\label{restrN1}
1 - |T|^2 \ll
\frac{1}{N}\,.
\end{equation}
It ensures that the oscillations of the Wigner function, which are
typically observed for a Fock state, are resolved.
Hence, the largest teleportation distance that is possible
scales (for large $N$) as
\begin{equation}
\label{restrN2}
l_{\rm T} \sim \frac{l_{\rm A}}{N}\,.
\end{equation}


\subsection{Discussion}
\label{sec3.5}

\begin{figure}[htb]
\vspace*{2ex}
\begin{center}
\mbox{\psfig{file=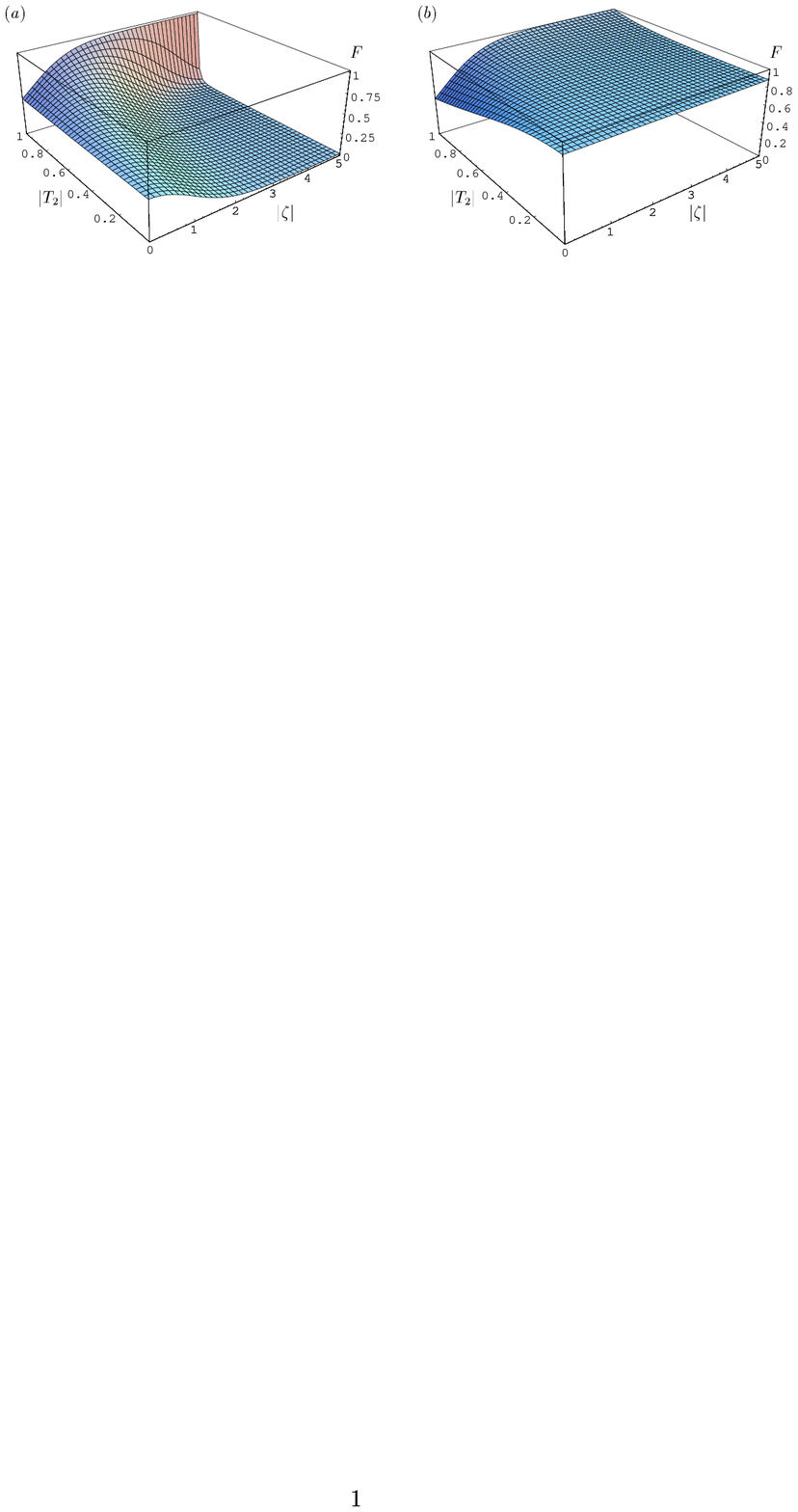,width=\textwidth}}
\end{center}
\vspace*{-4ex}
\caption{
\label{f1_tel}
The fidelity of teleportation of a squeezed vacuum state
\mbox{($\zeta_0$ $\!=$ $\!0.5$)} is shown as a function of $|\zeta|$
and $|T_2|$ (\mbox{$|T_1|$ $\!=$ $\!1$}, 
\mbox{$\tilde{\varphi}$ $\!=$ $\!0$}) for the displacement
$(a)$ \mbox{$\Delta(\gamma')$ $\!=$ $\!\gamma'$} and
$(b)$ \mbox{$\Delta(\gamma')$ $\!=$ $\!|T_2/T_1|\gamma'$}  
[Eqs.~(\protect\ref{asdis}) and (\protect\ref{lambda})].
}
\end{figure}

Whereas for perfect teleportation, i.e., $|T_1|$ $\!=$ $\!|T_2|$
$\!=$ $\!1$, Bob has to perform a displacement $\Delta(\gamma')$
$\!=$ $\!e^{i\varphi}\gamma'$
[Eq.~(\ref{asdis}) for \mbox{$\lambda$ $\!=$ $\!1$}],
which does not depend on the
position of the source of the TMSV, the situation drastically
changes for nonperfect teleportation. The effect is clearly seen 
from a comparison of Fig.~\ref{f1_tel}(a) with Fig.~\ref{f1_tel}$(b)$.
In the two figures, the fidelity for teleporting
a squeezed vacuum state is shown as a function of the
squeezing parameter $|\zeta|$ of the TMSV and the transmission
coefficient $|T_2|$ for the case when the source of the TMSV is
in Alice's hand, i.e., $|T_1|$ $\!=$ $\!1$.
Figure \ref{f1_tel}$(a)$ shows the result that is obtained
for \mbox{$\Delta(\gamma')$ $\!=$ $\!\gamma'$}. 
It is seen that when $|T_2|$ is not close to unity, then the
fidelity reduces, with increasing $|\zeta|$, below the classical
level (realized for \mbox{$|\zeta|$ $\!=$ $\!0$}). In contrast,
the displacement $\Delta(\gamma')$ $\!=$ $\!e^{i\tilde{\varphi}}
\lambda \gamma'$ with $\lambda$ from Eq.~(\ref{lambda}) ensures
that the fidelity exceeds the
classical level [Fig.~\ref{f1_tel}$(b)$].  

At this point the question may arise of whether the
choice of $\lambda$ according to Eq.~(\ref{lambda}) is the best one
or not. For example, from inspection of Eq.~(\ref{2.19}) it could
possibly be expected that $\lambda$ $\!=$ $\!C_2/|S|$ be also a 
a good choice. To answer the question, we note that in
the formulas for the teleported quantum state and the fidelity
$\lambda$ can be regarded
as being an arbitrary (positive) parameter that must not necessarily
be given by Eq.~(\ref{lambda}). Hence for chosen signal state
and given value of $|\zeta|$, the value of $\lambda$ (and thus
the value of the displacement) that maximizes the teleportation
fidelity can be determined. Examples of the fidelity (as a function
of $|\zeta|)$ that can be realized in this way are shown in
\begin{figure}[htb]
\vspace*{3ex}
\begin{center}
\mbox{\psfig{file=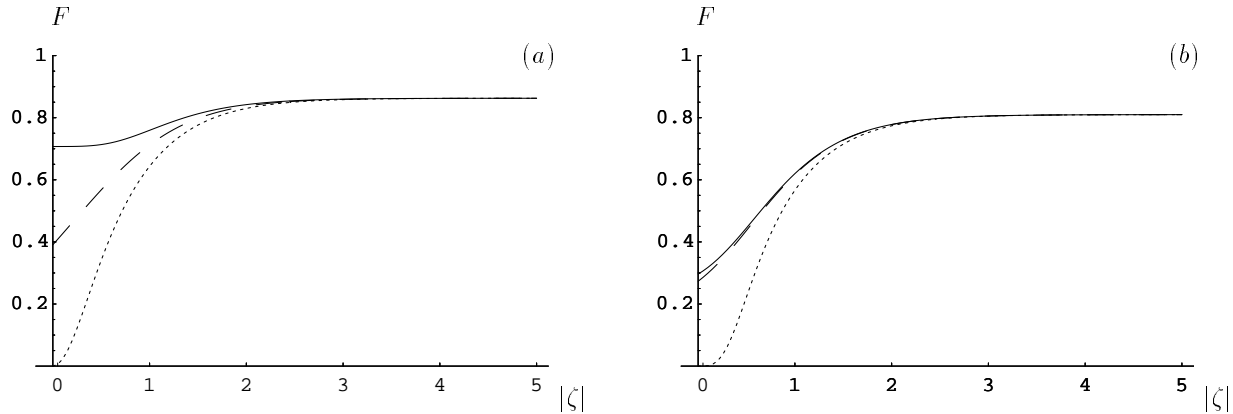,width=\textwidth}}
\end{center}
\vspace*{-4ex}
\caption{
\label{f2_tel}
The fidelity of teleportation of $(a)$ a squeezed vacuum state
(\mbox{$\zeta_0$ $\!=$ $\!0.88$}, i.e., \mbox{$\bar{n}$ $\!\approx$ $\!1$})
and $(b)$ a single-photon Fock state (\mbox{$N$ $\!=$ $\!1$})
is shown as a function of $|\zeta|$ (\mbox{$|T_1|$ $\!=$ $\!1$},
\mbox{$|T_2|$ $\!=$ $\!0.9$}, \mbox{$\tilde{\varphi}$ $\!=$ $\!0$}).
The parameter $\lambda$ in the displacement
\mbox{$\Delta(\gamma')$ $\!=$ $\!\lambda\gamma'$} is chosen such
that maximum fidelity is realized. For comparison, the fidelities
that are realized for \mbox{$\lambda$ $\!=$ $\!|T_2/T_1|$} (dashed line)
and \mbox{$\lambda$ $\!=$ $\!C_2/|S|$} (dotted line) are shown.
}
\end{figure}
Fig.~\ref{f2_tel} for teleporting squeezed and number
states according to Eqs.~(\ref{3.11}) and (\ref{fidfock}),
respecticvely. The figure reveals that for not too small
values of $|\zeta|$, that is, in the proper teleportation regime,
the state-{\em independent} choice of $\lambda$ according to
Eq.~(\ref{lambda}) is indeed the best one. 

\begin{figure}[htb]
\vspace*{2ex}
\begin{center}
\mbox{\psfig{file=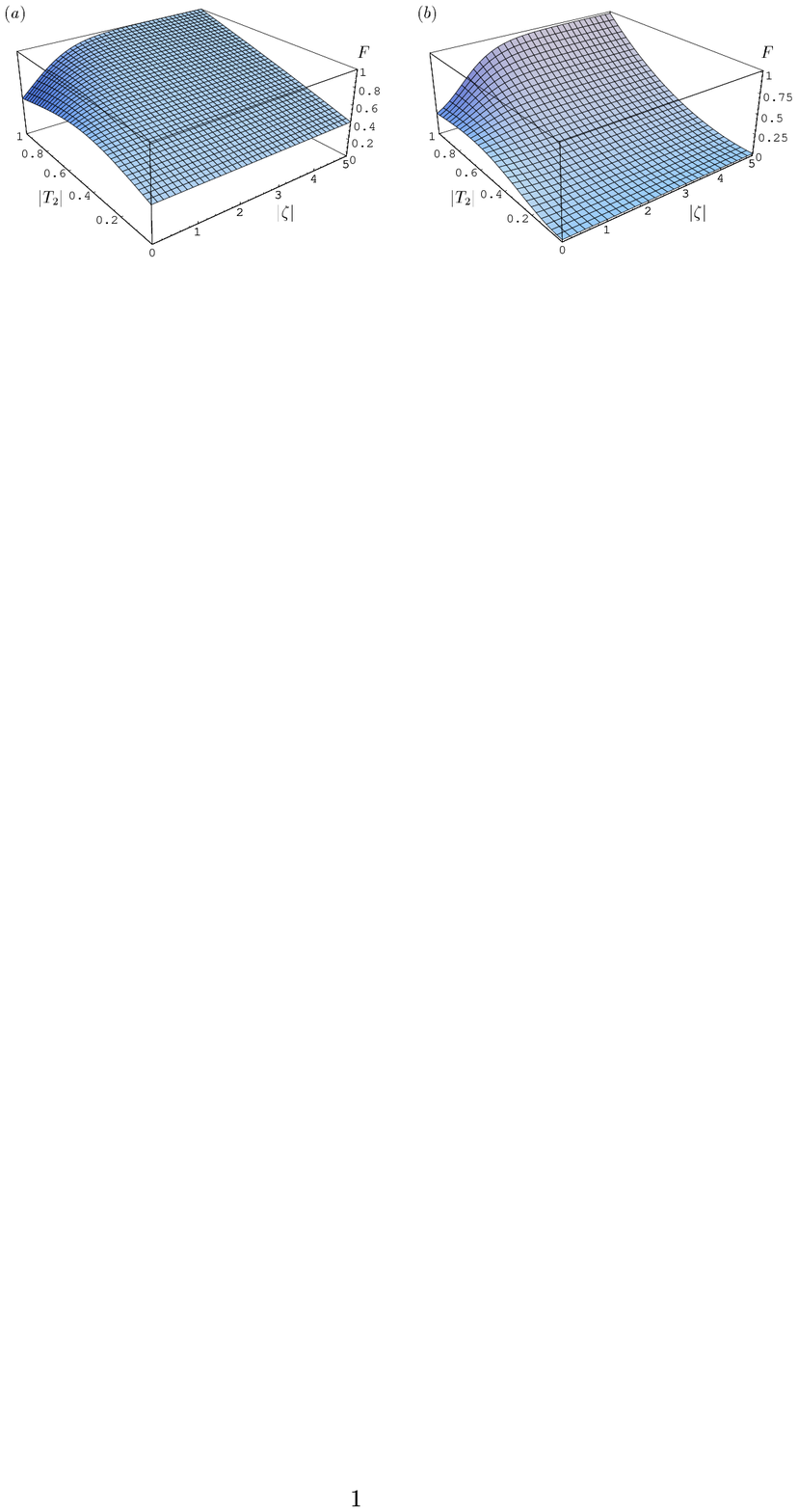,width=\textwidth}}
\end{center}
\vspace*{-4ex}
\caption{
\label{f3_tel}
The fidelity of teleportation of $(a)$ a squeezed coherent state 
(\mbox{$\zeta_0$ $\!=$ $\!0.5$}, \mbox{$\alpha_0$ $\!\approx$ $\!0.7$},
i.e., \mbox{$\bar{n}$ $\!\approx$ $\!1$})
and $(b)$ a single-photon Fock state (\mbox{$N$ $\!=$ $\!1$})
is shown as a function of $|\zeta|$ and $|T_2|$
(\mbox{$|T_1|$ $\!=$ $\!1$}, \mbox{$\tilde{\varphi}$ $\!=$ $\!0$},
\mbox{$\lambda$ $\!=$ $\!|T_2/T_1|$}).
}
\end{figure}

Figure \ref{f3_tel} illustrates the dependence of the teleportation
fidelity on the squeezing parameter $|\zeta|$ of the TMSV and the
transmission coefficient $|T_2|$ ($|T_1|$ $\!=$ $\!1$) 
for squeezed and number states.
It is seen that with increasing value of $|\zeta|$ the
fidelity is rapidly saturated below unity, because of absorption.
Even if the TMSV were infinitely entangled, the fidelity would
be noticeably smaller than unity in practice. 
Only when $|T_2|$ is very close to unity, the fidelity
substantially exceeds the classical level and becomes close
to unity. Note that the classical level is much smaller for
number states than for squeezed states.
Hence, it is principally impossible to
realize quantum teleportation over distances that are
comparable with those of classical channels. The result
is not unexpected, because the scheme is based on a strongly
squeezed TMSV, which corresponds to an entangled {\em macroscopic}
(at least {\em mesoscopic}) quantum state. As shown in
Section \ref{sec:distance}, such a state decays very rapidly,
so that the potencies inherent in it cannot be used in praxis.     

\begin{figure}[htb]
\vspace*{2ex}
\begin{center}
\mbox{\psfig{file=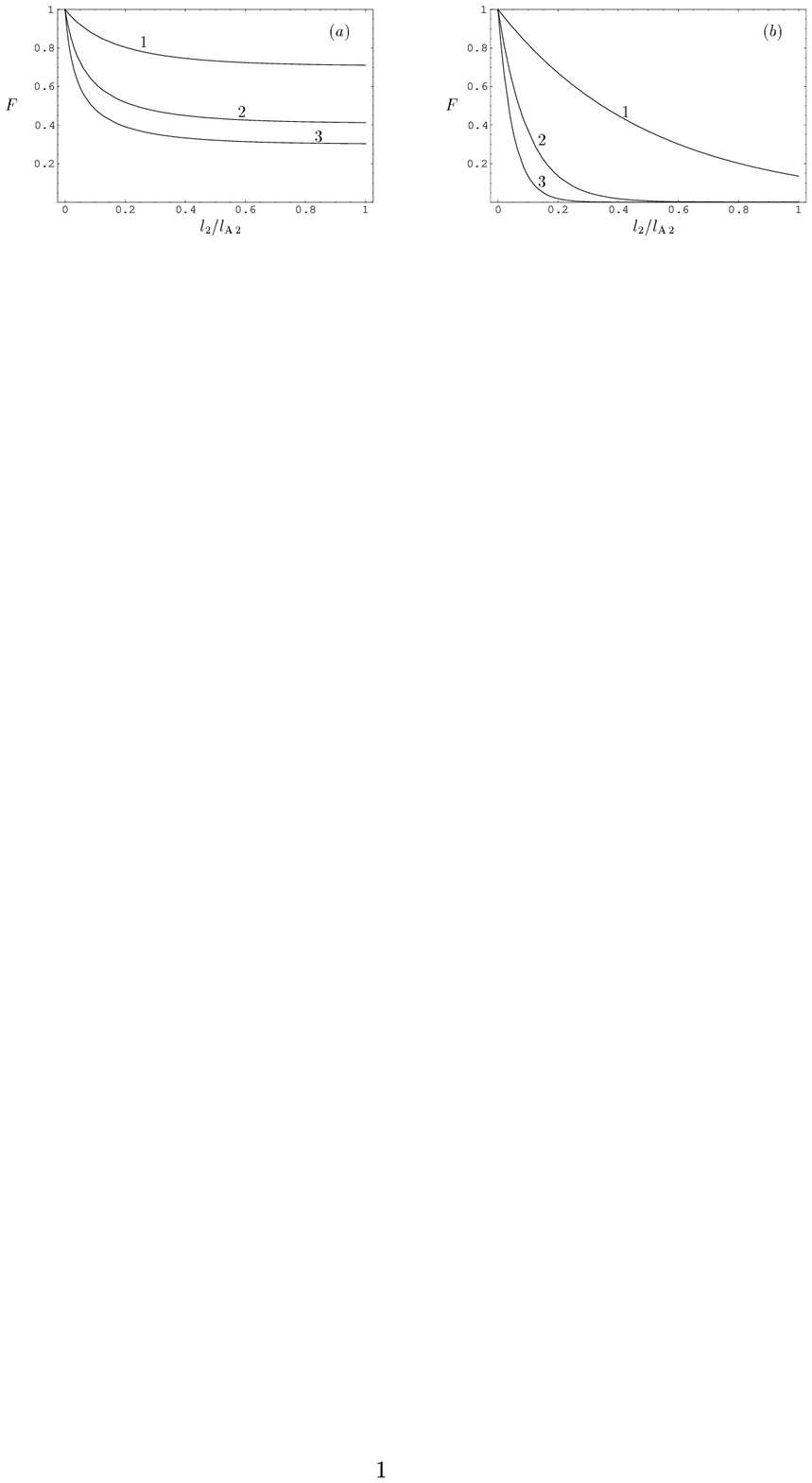,width=\textwidth}}
\end{center}
\vspace*{-4ex}
\caption{
\label{f4_tel}
The fidelity of teleportation of $(a)$ squeezed vacuum states
(curve $1$: \mbox{$\zeta_0$ $\!=$ $\!0.88$}, i.e.,
\mbox{$\bar{n}$ $\!\approx$ $\!1$};
curve $2$: \mbox{$\zeta_0$ $\!=$ $\!1.54$}, i.e.,
\mbox{$\bar{n}$ $\!\approx$ $\!5$};
curve $3$: \mbox{$\zeta_0$ $\!=$ $\!1.87$}, i.e.,
\mbox{$\bar{n}$ $\!\approx$ $\!10$})
and $(b)$ Fock states
(curve $1$: \mbox{$N$ $\!=$ $\!1$};
curve $2$: \mbox{$N$ $\!=$ $\!5$};
curve $3$: \mbox{$N$ $\!=$ $\!10$})
is shown as a function of the transmission length $l_2$
(\mbox{$|T_1|$ $\!=$ $\!1$}, \mbox{$\tilde{\varphi}$ $\!=$ $\!0$},
\mbox{$\lambda$ $\!=$ $\!|T_2/T_1|$}).
}
\end{figure}
In order to illustrate the ultimate limits in more detail,
we have plotted in Fig.~\ref{f4_tel} the dependence of the
teleportation fidelity on the transmission length $l_2$
(\mbox{$l_1$ $\!=$ $\!0$}) for squeezed and number
states, assuming an infinitely squeezed TMSV. It is seen
that with increasing transmission length the fidelity very
rapidly decreases, and it approaches the classical level on a
length scale that is much shorter than the absorption length. 
In particular, the distance over which a squeezed state can really
be teleported drastically decreases with increasing squeezing.
The same effect is observed for number states
when the number of photons increases. In other words,
for chosen distance, the amount of information that can   
be transferred quantum mechanically is limited,
so that essential information about the quantum state that is
desired to be teleported is lost. Obviously, this limitation
reflects the effect of saturation of entanglement, as it is
illustrated in Fig.~\ref{laenge}.    

\begin{figure}[htb]
\vspace*{4ex}
\begin{center}
\mbox{\psfig{file=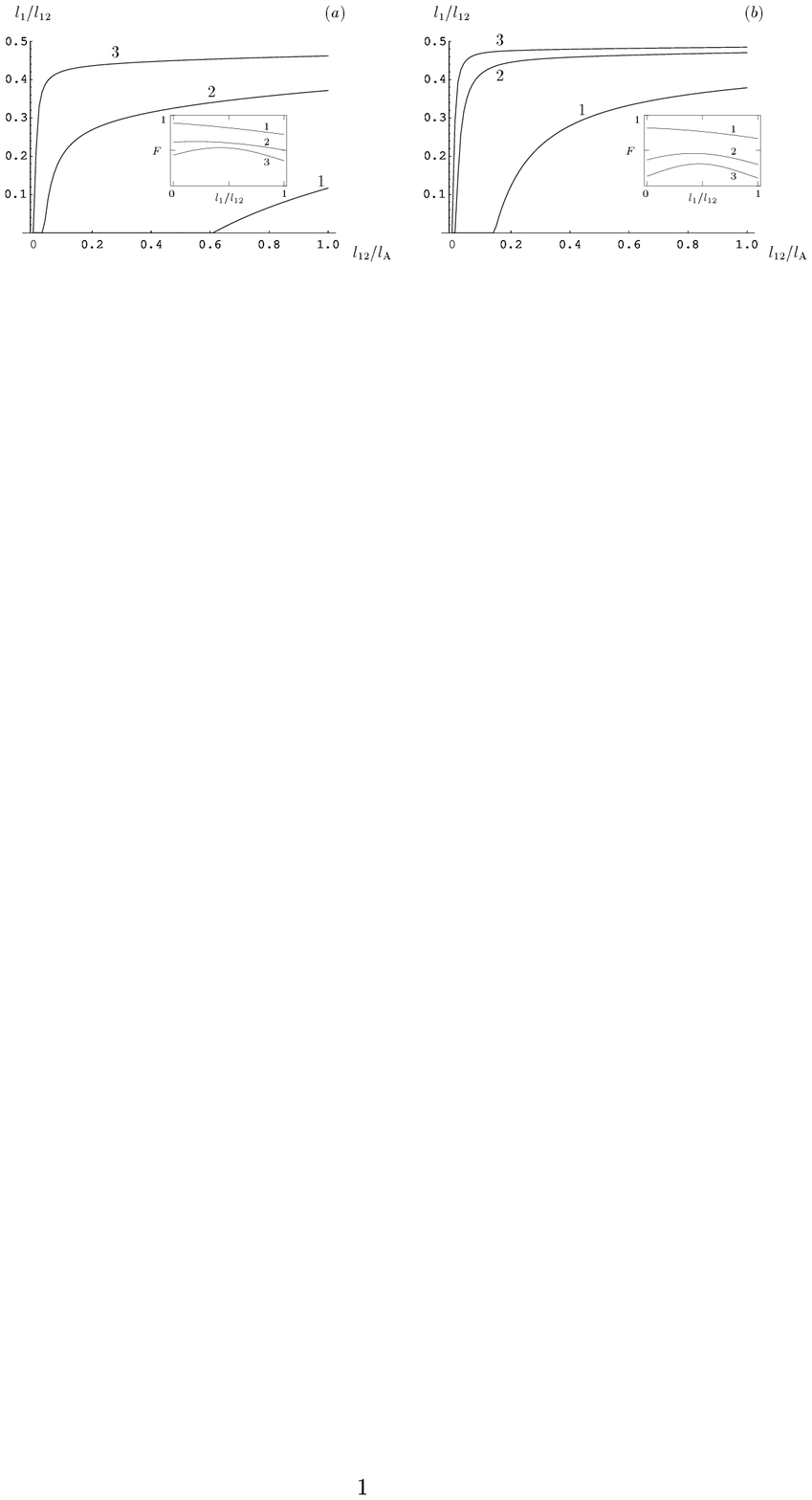,width=\textwidth}}
\end{center}
\vspace*{-4ex}
\caption{
\label{f5_tel}
The optimal distance $l_1$ from Alice to the position of an
infinitely squeezed TMSV, for which maximum teleportation fidelity
is realized, is shown as a function of the teleportation
distance \mbox{$l_{12}$
$\!=$ $\!l_1$ $\!+$ $\!l_2$}
(\mbox{$\tilde{\varphi}$ $\!=$ $\!0$},
\mbox{$\lambda$ $\!=$ $\!|T_2/T_1|$})
for $(a)$ squeezed states
(curve $1$: \mbox{$\zeta_0$ $\!=$ $\!0.78$},
\mbox{$\alpha_0$ $\!=$ $\!0.5$}, i.e.,
\mbox{$\bar{n}$ $\!\approx$ $\!1$};
curve $2$: \mbox{$\zeta_0$ $\!=$ $\!1.44$},
\mbox{$\alpha_0$ $\!=$ $\!1$}, i.e.,
\mbox{$\bar{n}$ $\!\approx$ $\!5$};
curve $3$: \mbox{$\zeta_0$ $\!=$ $\!1.63$},
\mbox{$\alpha_0$ $\!=$ $\!2$}, i.e.,
\mbox{$\bar{n}$ $\!\approx$ $\!10$})
and $(b)$ Fock states
(curve $1$: \mbox{$N$ $\!=$ $\!1$};
curve $2$: \mbox{$N$ $\!=$ $\!5$};
curve $3$: \mbox{$N$ $\!=$ $\!10$}). The insets show the
dependence of the fidelity on the position of the source of
the TMSV for \mbox{$l_{12}/l_{\rm A}$ $\!=$ $\!0.1$}. 
}
\end{figure}
So far we have considered the extremely asymetrical equipment
where the source of the TMSV
is in Alice' hand (\mbox{$|T_1|$ $\!=$ $\!1$}, i.e.,
\mbox{$|l_1|$ $\!=$ $\!0$}).
Whereas for perfect teleportation the source of the
TMSV can be placed anywhere, in praxis the teleportation fidelity
sensitively depends on the position of the source of
the TMSV. In Fig.~\ref{f5_tel}, examples of the optimal distance
$l_1$ from Alice to the source of an infinitely squeezed TMSV
(i.e., the distance for which maximum fidelity is realized) is shown,
again for squeezed and number states, as a function of the
distance \mbox{$l_{12}$ $\!=$ $\!l_1$ $\!+$ $\!l_2$} between Alice
and Bob (i.e., the transmission length). It is seen that the optimal
position of the source of the TMSV is state-dependent, and it is
always nearer to Alice than to Bob (\mbox{$0$ $\!\le$ $\!l_1$
$\!<$ $\!0.5\,l_{12}$}). With increasing value of $l_{12}$
the value of $l_1$ approaches $0.5\,l_{12}$, and one could thus
think that a symmetrical equipment would be the best one.
Unfortunately, this is not the case, because the transmission
lengths are essentially too large for true quantum teleportation.
What were (optimally) observed would be the classical level at best.    


\section{Summary and conclusions}
\label{sec4}

When entangled light is transmitted through optical devices, losses
always give rise to entanglement degradation. In particular, after
propagation of the two modes of a TMSV through fibers the available
entanglement can be drastically reduced. Unfortunately, quantifying
entanglement of mixed states in infinite-dimensional Hilbert
spaces has been close to impossible. Therefore, estimates and upper
bounds for the entanglement content have been developed.

In order to
quantify the entanglement degradation of a TMSV more precisely, 
we have considered the distance of the output Gaussian state to the
set of separable Gaussian states measured by the relative entropy.
It has the advantage that separable states obviously correspond
to zero distance. Although one has yet no proof that there does
not exist a non-Gaussian separable state which is closer to the Gaussian
state under consideration than the closest separable Gaussian
state, one has good reason to think that it is even an
entanglement measure. In any case, it is a much better bound
than the one obtained by convexity. In particular, it clearly
demonstrates the drastic decrease of entanglement of the output
state with increasing entanglement of the input state.
Moreover, one observes saturation of entanglement transfer;
that is, the amount of entanglement that can maximally
be contained in the output state is solely determined by
the transmission length and does not depend on the
amount of entanglement contained in the input state. 

In continuous-variable quantum teleportation it is commonly
assumed that Alice and Bob share an infinitely squeezed
TMSV. Since the TMSV state as an effectively
macroscopic (at least mesoscopic) entangled quantum state
which rapidly decays, proper quantum teleportation can be
expected to be possible only over distances that are much more
shorter than the (classical) absorption length.
Our analysis shows that this is indeed the case.
The strong entanglement degradation
dramatically limits the amount of information that can
be transferred quantum mechanically over longer
distances.

Because of this limitation, quantum teleportation
becomes state-dependent, that is, without additional knowledge
of the state that is desired to be teleported over some
finite distance it is principally impossible to decide whether
the teleported state is sufficiently close to the original state.
It is worth noting that both the coherent displacement that
must be performed by Bob and the position of the source of
the TMSV should not be chosen independently of the fiber lengths.
In particular, an asymmetrical equipment, where the source of
the TMSV is placed nearer to Alice than to Bob, is  suited
for realizing the largest possible teleportation fidelity
and not a symmetrical one. 
     
Throughout this paper we have restricted our attention to
(quasi-)monochromatic fields. Using wave packets, the
ultimate limits of quantum teleportation are not only determined
by absorption but also  by dispersion. Due to dispersion, the
two wave packets unavoidably change their forms during propagation
over longer distances, and the problem of mode mismatching
in Alice's homodyne measuerment and Bob's coherent displacement 
appears. The corresponding quantum efficiencies
diminish, and hence the width of the Gaussian with which 
the Wigner function of the original quantum state is convolved
is effectively increased. As a result, the teleportation    
fidelity is reduced. It can be expected that the effect
sensitively depends on the position of the source of the TMSV.
In order to understand the details, a separate analysis is
required, which will be given elsewhere.
 

\section*{Acknowledgements}
This work was supported by the Deutsche Forschungsgemeinschaft.


\end{document}